\RequirePackage{ifpdf}
\documentclass[hyper]{JHEP3}
\usepackage{graphicx}
\usepackage{epsfig}
\usepackage{multicol}
\usepackage{bbm}
\usepackage{amsmath}
\usepackage{cite}
\usepackage{lscape}
\usepackage{multirow}
\usepackage{amscd} 
\usepackage{slashed}
\usepackage{subfigure}
\usepackage{booktabs}

\DeclareMathSymbol{\Gamma}{\mathalpha}{letters}{"00}
\DeclareMathSymbol{\Delta}{\mathalpha}{letters}{"01}
\DeclareMathSymbol{\Theta}{\mathalpha}{letters}{"02}
\DeclareMathSymbol{\Lambda}{\mathalpha}{letters}{"03}
\DeclareMathSymbol{\Xi}{\mathalpha}{letters}{"04}
\DeclareMathSymbol{\Pi}{\mathalpha}{letters}{"05}
\DeclareMathSymbol{\Sigma}{\mathalpha}{letters}{"06}
\DeclareMathSymbol{\Upsilon}{\mathalpha}{letters}{"07}
\DeclareMathSymbol{\Phi}{\mathalpha}{letters}{"08}
\DeclareMathSymbol{\Psi}{\mathalpha}{letters}{"09}
\DeclareMathSymbol{\Omega}{\mathalpha}{letters}{"0A}
\DeclareMathSymbol{\varGamma}{\mathalpha}{operators}{"00}
\DeclareMathSymbol{\varDelta}{\mathalpha}{operators}{"01}
\DeclareMathSymbol{\varTheta}{\mathalpha}{operators}{"02}
\DeclareMathSymbol{\varLambda}{\mathalpha}{operators}{"03}
\DeclareMathSymbol{\varXi}{\mathalpha}{operators}{"04}
\DeclareMathSymbol{\varPi}{\mathalpha}{operators}{"05}
\DeclareMathSymbol{\varSigma}{\mathalpha}{operators}{"06}
\DeclareMathSymbol{\varUpsilon}{\mathalpha}{operators}{"07}
\DeclareMathSymbol{\varPhi}{\mathalpha}{operators}{"08}
\DeclareMathSymbol{\varPsi}{\mathalpha}{operators}{"09}
\DeclareMathSymbol{\varOmega}{\mathalpha}{operators}{"0A}

\renewcommand{\vec}[1]{\boldsymbol{#1}}
\newcommand{\gev}{\text{\,GeV}}
\newcommand{\tev}{\text{\,TeV}}

\def\bi{\begin{itemize}}
\def\ei{\end{itemize}}
\newcommand{\msq}{{m_{\widetilde{q}}}}
\newcommand{\mgo}{{m_{\widetilde{g}}}}
\newcommand{\sq}{{\widetilde{q}}}
\newcommand{\go}{{\widetilde{g}}}
\renewcommand{\L}{{\text{L}}}
\newcommand{\R}{{\text{R}}}
\newcommand{\T}{{\text{T}}}
\newcommand{\s}[1]{\widetilde{#1}}
\newcommand{\pb}{\ensuremath{\,\textnormal{pb}}}
\newcommand{\fb}{\ensuremath{\,\textnormal{fb}}}

\newcommand{\Tinv}{\text{T2}_\infty}
\newcommand{\Tmgo}{\text{T2}_\mgo}
\newcommand{\aT}{\alpha_{\text{T}}}
\newcommand{\mne}{m_{\s\chi^0_1}}

\allowdisplaybreaks

\title{Constraining supersymmetry at the LHC with simplified models for squark production}

\author{
Lisa Edelh\"auser${}^a$, Jan Heisig${}^a$, Michael Kr\"amer${}^{a,b}$, Lennart Oymanns${}^a$, and Jory Sonneveld${}^a$\\
${}^a$Institute for Theoretical Particle Physics and Cosmology, RWTH Aachen University,\\ \hspace*{1.75mm}52056 Aachen, Germany

\vspace*{1mm}

${}^b$SLAC National Accelerator Laboratory, Stanford University, Stanford, CA 94025, USA}

\abstract{An important tool for interpreting LHC searches for 
new physics are simplified models. They are characterized by a 
small number of parameters and thus often rely on a 
simplified description of particle production and decay dynamics.
Considering the production of squarks of the first two generations
we compare the interpretation of current LHC searches for hadronic jets plus missing energy signatures
within simplified models with the interpretation within a complete supersymmetric model.  Although 
we find sizable differences in the signal efficiencies, in particular for large supersymmetric particle masses, the differences between the mass limits derived from a simplified model and 
from the complete supersymmetric model are moderate given the current LHC sensitivity. 
We conclude that simplified models provide a 
reliable tool to interpret the current hadronic jets plus missing energy searches at the LHC in a 
more model-independent way.}


\preprint{
SLAC-PUB-16114\\
TTK-14-26} 

\begin{document}
%%%%%%%%%%%%%%%%%%%%%%%%%%%%%%%%%%%%%%%%%%%%%%%%%%%%%%%

%===================================================================
\section{Introduction}
%===================================================================

The quest for covering a large number of scenarios beyond the Standard Model (BSM)
with current searches for new physics at the LHC requires efficient
ways of matching experimental results with theory predictions.
A particularly successful approach is the utilization of simplified models
\cite{ArkaniHamed:2007fw,Alwall:2008ag,Alves:2011wf,sidewalk}. Recently developed program 
packages~\cite{Kraml:2013mwa,Papucci:2014rja} provide a convenient framework to employ simplified models for testing BSM
theories  at the LHC\@.
Common to the experimental limits thus obtained is that the simplified models used for data interpretation are characterized by a small number of new 
particles and a simplified description of particle production and decay. 
The underlying assumption for this treatment is that the more model-specific details of the production and decay 
dynamics have little influence on the signal efficiencies. In this study we question the validity of this assumption. 
We focus on light flavor squark production in the minimal supersymmetric model with $R$-parity conservation. 
In supersymmetric models, squark production proceeds also through the exchange of a gluino in the $t$-channel and 
thus includes various processes with left- and right-chiral 
squarks and anti-squarks in the final state, $pp\to \s{q}_i^{}\s{q}_j^*$ and $pp\to \s{q}_i\s{q}_j$, with the chirality 
$i,j = \text{L},\text{R}$. In the simplified models adopted by the ATLAS and CMS collaborations, however, the usual choice is 
to decouple the gluino, see, e.g., \cite{Aad:2014wea, Chatrchyan:2013lya,Chatrchyan:2014lfa}. Consequently, the only contributing production mode is $pp\to\sq_i^{}\sq_i^*$.
As $pp\to\sq_i^{}\sq_i^*$ is in general not the dominant production channel, 
it is an important question whether the signal efficiencies -- and hence the resulting exclusion limits --
derived from this production mode are applicable to the more general case of supersymmetric models. 

In this paper we compare the efficiencies for squark production and the exclusion limits for squark 
masses obtained by using a simplified model with those obtained in the complete supersymmetric model. 
We focus on two representative 
all-hadronic analyses performed by CMS: 
one based on the discriminating variable $\alpha_\T$~\cite{Chatrchyan:2013lya}, and one based on missing transverse 
energy, denoted by $\slashed{H}_\T$~\cite{Chatrchyan:2014lfa}.
As in the experimental analyses, we assume a direct decay of the squark into the 
neutralino. The respective topology is often referred to as T2~\cite{Chatrchyan:2013sza}. The simplified model with a decoupled gluino is 
denoted by $\Tinv$ in the following, while the complete supersymmetric model with a finite gluino mass is denoted by $\Tmgo$. 
The simplified model commonly adopted by ATLAS and CMS for the hadronic searches for squarks 
corresponds to $\Tinv$. 

We find very large deviations between the efficiencies for $\Tinv$ and $\Tmgo$ in certain regions of parameter space, in particular for large squark masses. 
However, we show that for the parameter region relevant for the current 
exclusion limits, the differences turn out to be moderate, and the exclusion limits
obtained from applying the efficiencies for $\Tinv$ are close to the ones for $\Tmgo$.
Comparing the differences to the theoretical uncertainties of the cross section normalization, 
we find that both are of the same size for large regions in the considered parameter space.
In addition, we compare the two analyses based on $\alpha_\T$ and $\slashed{H}_\T$ and show
that  the $\alpha_\T$ variable is less sensitive to the difference in the production dynamics and 
modes. Furthermore, whereas in the $\slashed{H}_\T$ analysis the $\Tinv$ efficiencies lead to an overestimation of the
exclusion limits, in the $\alpha_\T$ analysis $\Tinv$ yields conservative limits.

This paper is structured as follows. In section \ref{sec:production} we review the different squark production 
processes and their dependence on the gluino mass. In section \ref{sec:analysis} we introduce the two CMS searches
and the parameter scan. The results for the comparison of efficiencies and squark mass limits are presented in section \ref{sec:results}. We conclude in section~\ref{sec:conclusion}.

%===================================================================
\section{Production processes of squarks at the LHC}
%===================================================================
\label{sec:production}

\begin{figure}[htp]
\centering
\setlength{\unitlength}{1\textwidth}
\begin{picture}(0.9,0.34)
 \put(-0.024,0.18){ % upper
 \put(-0.024,0.06){ a)}
~~~\includegraphics[width=3.2cm]{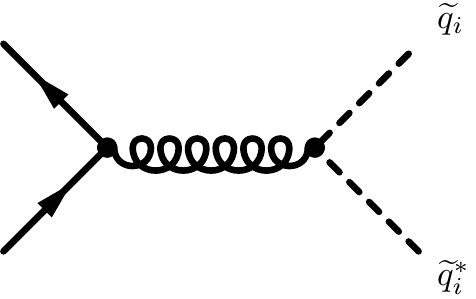}~~
\includegraphics[width=3.2cm]{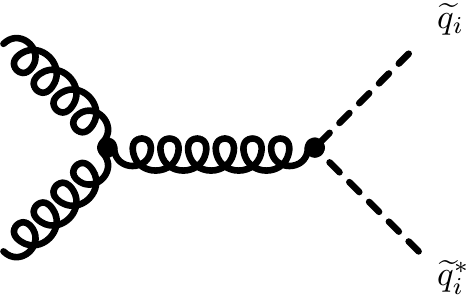}~~
\includegraphics[width=3.2cm]{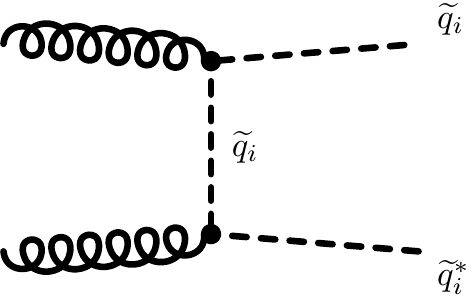}~~
\includegraphics[width=3.2cm]{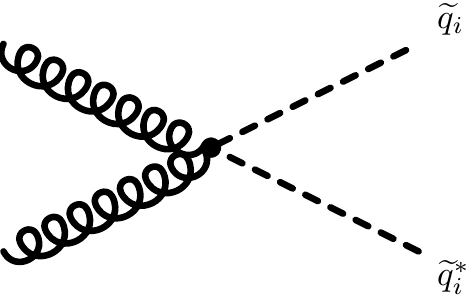}~~
}
 \put(-0.024,0.0){% lower
 \put(-0.024,0.06){ b)}
~~~\includegraphics[width=3.2cm]{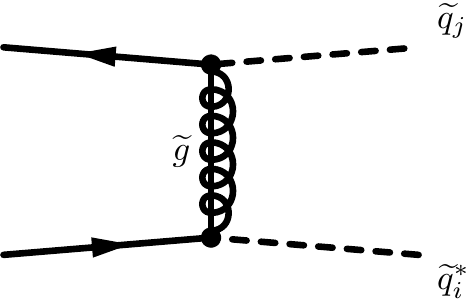}~~~~~~
\includegraphics[width=3.2cm]{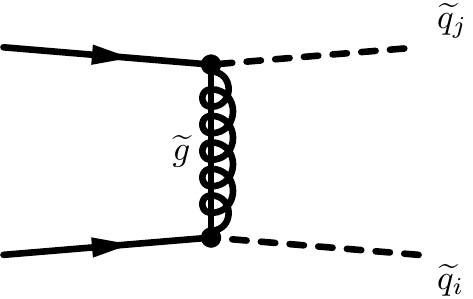}
}
\end{picture}
\caption{a)~Production process $pp\rightarrow \widetilde{q}_i^{}\widetilde{q}^*_i$ with $i=\text{L},\text{R}$ with a decoupled gluino. For $i=\text{L}$, these graphs contribute to $\Tinv$. b)~Diagrams for $pp\rightarrow \widetilde{q}_i^{}\widetilde{q}^*_j$  and $pp\rightarrow \widetilde{q}_i^{}\widetilde{q}_j^{}$, with $i,j=\text{L},\text{R}$. These are present if the gluino is not decoupled. Thus both types of diagrams a) and b) contribute to $\Tmgo$.}
\label{fig:sqsqbar_production}
\end{figure}

The production of strongly interacting sparticles is an important production channel for
probing supersymmetry at the LHC.
The strength and the kinematic distribution of squark (anti-)squark production
not only depend on the mass spectrum of the 
squarks, but also on the 
gluino mass $\mgo$, since all squark production processes -- except for the case $pp\to \s{q}_i^{}\s{q}_i^*$, $i=\text{L},\text{R}$ --
require a $t$-channel gluino exchange 
(cf. the diagrams in figure \ref{fig:sqsqbar_production}).

%=====================
%    \                                           |
%      \                                         |
%        \                                       |
%                                      \         |
%                                        \       |
%                                          \     |
%=====================

We first discuss the relative importance of different squark production processes as a function
of the mass ratio $\mgo/\msq$,
where we make the convenient assumption 
of a common mass $\msq$ for the squarks of the first and second generation.\footnote{%
Here we consider $\mgo$ and $\msq$ as free parameters defined at the TeV scale. Note, however, that a
large ratio $\mgo/\msq$ is inaccessible in models where the supersymmetry breaking is mediated at a high scale.
For a detailed discussion see, e.g., \cite{Jaeckel:2011wp}.}
The third generation 
is not considered here. While varying the ratio $\mgo/\msq$, we kept the total production 
cross section of squarks and gluinos, denoted by $\sigma_{\{\go\sq\}}$, fixed. This requirement determines 
$\msq$ and $\mgo$.
We show relative cross section contributions for $\sigma_{\{\go\sq\}}\simeq 1000\fb$ and $\sigma_{\{\go\sq\}}\simeq 10\fb$. These values
represent typical cross section upper limits from the 8\,TeV LHC null-search for
regions with very small and very high sensitivities, respectively. For $\mgo/\msq\gtrsim3$, this 
corresponds to squark masses of about $500\gev$ and $1\tev$, respectively.
We computed the total production cross section at NLO with \textsc{Prospino} \cite{Beenakker:1996ch}, while the individual
contributions were calculated at LO with \textsc{MadGraph}~5~\cite{Alwall:2011uj}\@.

%=====================
%    \                                           |
%      \                                         |
%        \                                       |
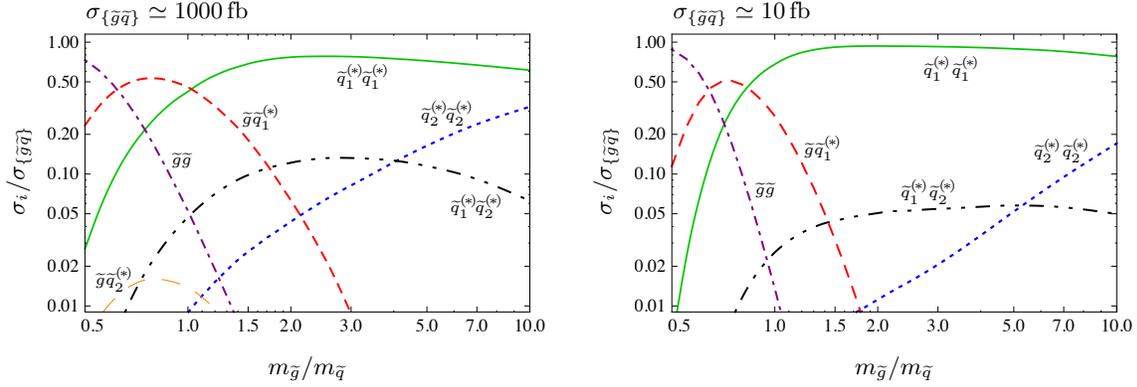
\begin{figure}[t]
\centering
\setlength{\unitlength}{1\textwidth}
\begin{picture}(0.893,0.354)
 \put(-0.024,0.0){ % left
  \put(0.0,0.025){\includegraphics[width=0.45\textwidth]{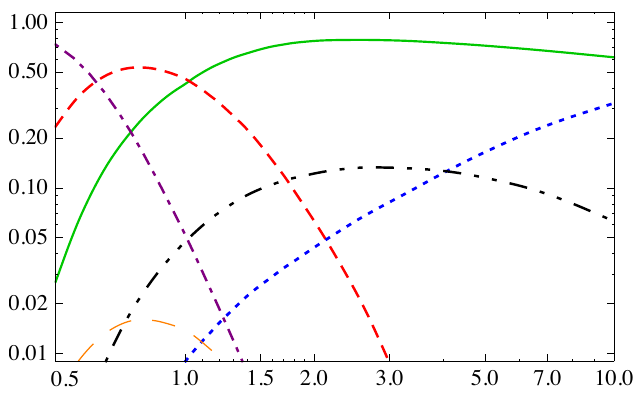}}
  \put(0.2,0.0){\footnotesize $\mgo/\msq$}
  \put(-0.03,0.14){\rotatebox{90}{\footnotesize $\sigma_i/\sigma_{\{\go\sq\}}$}}
  \put(0.26,0.256){\tiny $\sq_1^{(\!*\!)}\!\sq_1^{(\!*\!)}$}
  \put(0.177,0.22){\tiny $\go\sq_1^{(\!*\!)}$}
  \put(0.115,0.184){\tiny $\go\go$}
  \put(0.36,0.142){\tiny $\sq_1^{(\!*\!)}\!\sq_2^{(\!*\!)}$}
  \put(0.334,0.224){\tiny $\sq_2^{(\!*\!)}\!\sq_2^{(\!*\!)}$}
  \put(0.048,0.078){\tiny $\go\sq_2^{(\!*\!)}$}
  \put(0.04,0.313){\footnotesize $\sigma_{\{\go\sq\}}\simeq1000\fb$}
  }
 \put(0.492,0.0){ % right
  \put(0.0,0.025){\includegraphics[width=0.45\textwidth]{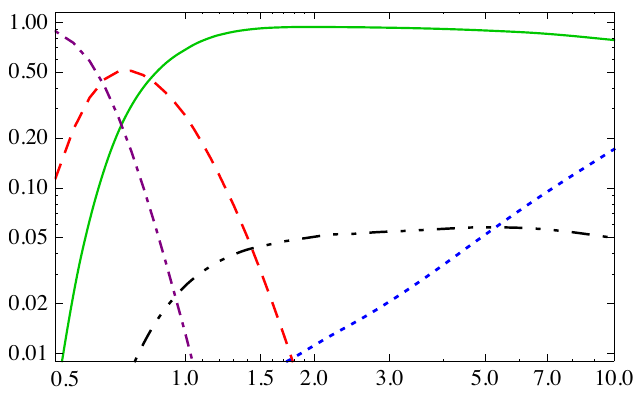}}
  \put(0.2,0.0){\footnotesize $\mgo/\msq$}
  \put(-0.03,0.14){\rotatebox{90}{\footnotesize $\sigma_i/\sigma_{\{\go\sq\}}$}}
  \put(0.26,0.263){\tiny $\sq_1^{(\!*\!)}\sq_1^{(\!*\!)}$}
  \put(0.155,0.193){\tiny $\go\sq_1^{(\!*\!)}$}
  \put(0.112,0.158){\tiny $\go\go$}
  \put(0.24,0.155){\tiny $\sq_1^{(\!*\!)}\sq_2^{(\!*\!)}$}
  \put(0.357,0.191){\tiny $\sq_2^{(\!*\!)}\sq_2^{(\!*\!)}$}
  \put(0.04,0.313){\footnotesize $\sigma_{\{\go\sq\}}\simeq10\fb$}
  }
\end{picture}
\caption{
Relative contributions to the production cross section of squarks and gluinos
as a function of the ratio $\mgo/\msq$ along iso-cross section curves for the $8\tev$ LHC\@. 
\emph{Left panel:}~For a total production cross section $\sigma_{\{\go\sq\}}\simeq 1000\fb$, i.e.,
small $\mgo$ and $\msq$. 
\emph{Right panel:}~For $\sigma_{\{\go\sq\}}\simeq 10\fb$, i.e.,
large $\mgo$ and $\msq$.  We take into account all production 
mechanisms of gluinos and of squarks of the first and second generation. These 
squarks are assumed to have a common mass $\msq$. Here we denote 
$\sq^{(*)}_1$, $\sq^{(*)}_2$ 
to be all first and second generation (anti)squarks not distinguishing between
left- and right superpartners. 
}
\label{fig:contri1}
\end{figure}
%                                      \         |
%                                        \       |
%                                          \     |
%=====================

%=====================
%    \                                           |
%      \                                         |
%        \                                       |
\begin{figure}[h!]
\centering
\setlength{\unitlength}{1\textwidth}
\begin{picture}(0.893,0.35)
 \put(-0.024,0){ % left
  \put(0.0,0.025){\includegraphics[width=0.45\textwidth]{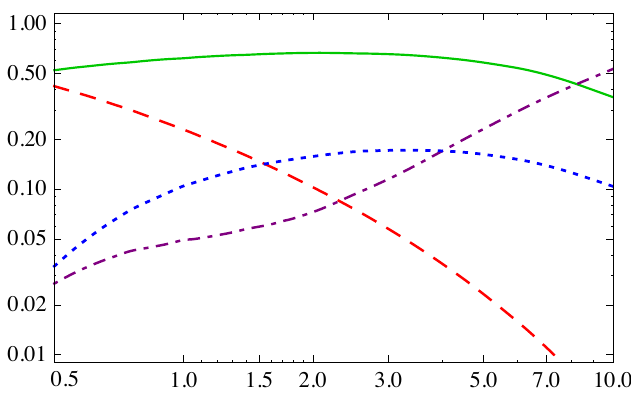}}
  \put(0.2,0.0){\footnotesize $\mgo/\msq$}
  \put(-0.03,0.14){\rotatebox{90}{\footnotesize $\sigma_i/\sigma_{\sq_1\sq_1}$}}
  \put(0.2,0.254){\tiny  $\sq_\L\sq_\L\!+\!\sq_\R\sq_\R$}
  \put(0.12,0.227){\tiny $\sq_\L\sq_\R$}
  \put(0.1,0.115){\tiny $\sq_\L^{}\sq_\L^*\!+\!\sq_\R^{}\sq_\R^*$}
  \put(0.33,0.163){\tiny $\sq_\L^{}\sq_\R^* \!+\! \sq_\R^{}\sq_\L^*$}
  \put(0.01,0.313){\footnotesize Subcontributions of $\sq_1^{(*)}\sq_1^{(*)}$, $\sigma_{\{\go\sq\}}\simeq1000\fb$}
  }
 \put(0.492,0){ % right
  \put(0.0,0.025){\includegraphics[width=0.45\textwidth]{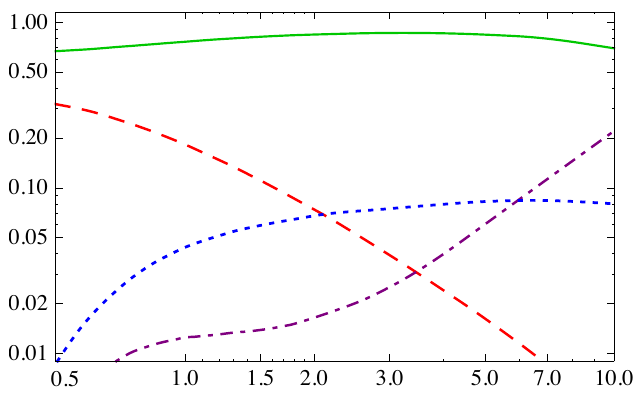}}
  \put(0.2,0.0){\footnotesize $\mgo/\msq$}
  \put(-0.03,0.14){\rotatebox{90}{\footnotesize $\sigma_i/\sigma_{\sq_1\sq_1}$}}
  \put(0.2,0.265){\tiny  $\sq_\L\sq_\L\!+\!\sq_\R\sq_\R$}
  \put(0.1,0.225){\tiny $\sq_\L\sq_\R$}
  \put(0.105,0.086){\tiny $\sq_\L^{}\sq_\L^*\!+\!\sq_\R^{}\sq_\R^*$}
  \put(0.05,0.148){\tiny $\sq_\L^{}\sq_\R^* \!+\! \sq_\R^{}\sq_\L^*$}
  \put(0.029,0.313){\footnotesize Subcontributions of $\sq_1^{(*)}\sq_1^{(*)}$, $\sigma_{\{\go\sq\}}\simeq10\fb$}
  }
\end{picture}
\caption{
Relative contributions to the production cross section of squarks of the first
generation at the $8\tev$ LHC\@.
These are the subcontributions of $\sq_1^{(*)}\sq_1^{(*)}$ in figure~\protect\ref{fig:contri1} 
 as a function of the ratio $\mgo/\msq$ along iso-cross section curves. 
 \emph{Left panel:}~For a total production cross section $\sigma_{\{\go\sq\}}\simeq 1\pb$, corresponding to  
small squark masses. 
\emph{Right panel:}~For $\sigma_{\{\go\sq\}}\simeq 10\fb$, corresponding to 
large squark masses.}
\label{fig:contri2}
\end{figure}
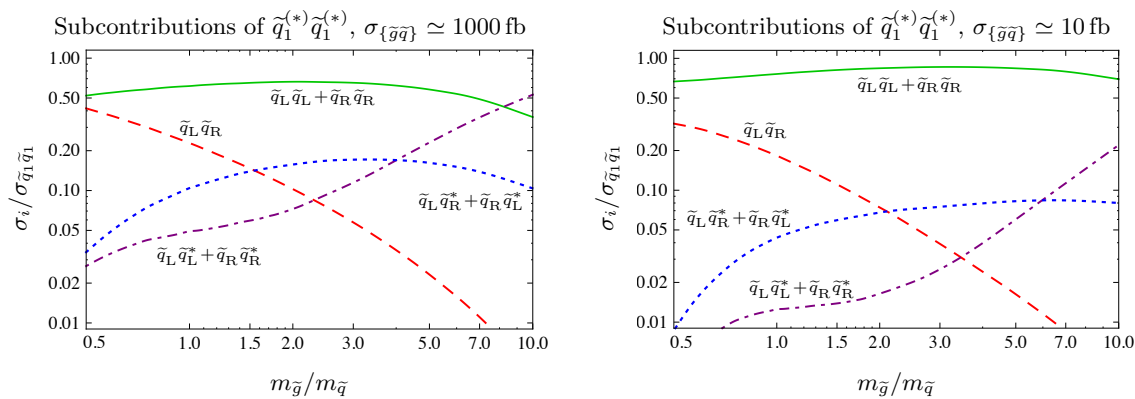
%                                      \         |
%                                        \       |
%                                          \     |
%=====================

In figure \ref{fig:contri1} we show the 
relative strength of all possible combinations of first generation squark production, $\sq_1^{(*)}$, second generation
squark production, $\sq_2^{(*)}$, and gluino production. In figure~\ref{fig:contri2}, the subcontributions to first generation squark pair
production, $\sq_1^{(*)}\sq_1^{(*)}$, are shown.  We summed processes that give equal contributions, such as 
$\sq_\L^{}\sq_\L^{}$ and $\sq_\R^{}\sq_\R^{}$. Subcontributions that are not displayed (antisquark pair
production, for instance) are below the range displayed here. 
From figures \ref{fig:contri1} and \ref{fig:contri2} we can draw several conclusions. 
First, the production of first generation squark pairs, $\sq_1^{(*)}\sq_1^{(*)}$, is the dominant production channel 
over a large range of  $\mgo/\msq$. This dominance is even more pronounced for $\sigma_{\{\go\sq\}}\simeq 10\fb$, i.e.,
for larger $\msq$. 
Second, among its subcontributions, the channel $\sq_\L^{}\sq_\L^{}+\sq_\R^{}\sq_\R^{}$
(with $\sq \equiv \sq_1=\s u, \s d\,$) is the dominant channel. Although for large $\mgo/\msq$ its contribution is 
suppressed through the gluino mass appearing in the $t$-channel propagator, the relative contributions
stays dominant up to $\mgo\simeq7\msq$ (for $\sigma_{\{\go\sq\}}\simeq 1000\fb$) and above 
$\mgo/\msq=10$ (for $\sigma_{\{\go\sq\}}\simeq 10\fb$). These effects are due
to the relatively large parton luminosities for $uu$ and $dd$ initial states 
when approaching large partonic center-of-mass energies in the hard scattering process. In contrast,
although equally enhanced through a higher parton luminosity, 
the relative contribution of $\sq_\L^{}\sq_\R^{}$ decreases rapidly with increasing $\mgo/\msq$. This is
because its cross section is suppressed by $1/\mgo^2$ compared to $\sq_\L^{}\sq_\L^{}$ and $\sq_\R^{}\sq_\R^{}$ 
in the limit of large $\mgo$.

The subcontributions to second generation squark pair production are completely 
dominated by $\sq_\L^{}\sq^*_\L$ and $\sq_\R^{}\sq^*_\R$. Its absolute contribution is 
dominated by the diagrams in the first row of figure~\ref{fig:sqsqbar_production},
which are independent of the gluino mass as well as independent of the squark 
flavor.

In summary, even for $\mgo/\msq$ as large as 10 the channels $\sq_\L^{}\sq^*_\L$, $\sq_\R^{}\sq^*_\R$
are not necessarily the dominant production mode of squarks, and it is crucial to take into
account other channels, in particular $\sq_\L^{}\sq_\L^{}$ and $\sq_\R^{}\sq_\R^{}$. These, in turn, yield different event
kinematics and thus, in principle, different signal efficiencies for the experimental searches.

%===================================================================
\section{Analyses and parameter scan}
\label{sec:analysis}
%===================================================================

We consider two representative all-hadronic analyses performed by the CMS collaboration using the 
$8\tev$ LHC data set. One analysis is based on the discriminating variable $\alpha_\T$~\cite{Chatrchyan:2013lya}, 
and one is based on missing transverse energy $\slashed{H}_\T$ \cite{Chatrchyan:2014lfa}.
Among other models, these searches were interpreted in terms of the simplified
model $\Tinv$, which consists of squark production via the process $pp\to\sq\sq^*\;(\mgo\to\infty)$,
followed by a squark decay into a neutralino $\s{\chi}^0_1$ (the lightest supersymmetric particle, LSP, in this model) and a quark with branching ratio 1.

%----------------------------------------------------------------------------------------------------------------
\subsection{All-hadronic analyses}
\label{sec:analyses}
%----------------------------------------------------------------------------------------------------------------

The CMS analysis \cite{Chatrchyan:2014lfa} is sensitive to the pair production 
of squarks (and gluinos) decaying into one (two) jets and an LSP. The search requires at least 
three jets, and the data is divided into several categories (3--5, 6--7, 8 jets). Consequently, the 
analysis is more sensitive to final states resulting from longer cascade decays of gluinos and 
squarks. Since we are interested in the pair production of squarks, we concentrate on the 3--5 
jet analysis. The search is based on the two variables

\begin{align}
H_\T&= \sum_{j\,\in\,\text{jets}} p^j_\T&& \text{for jets with} \quad p^j_\T>50 \gev\,,\; |\eta_j|<2.5\,, \\
\slashed{H}_\T& =\left|\slashed{\vec{H}}_\T\right|=\left|-\sum_{j\,\in\,\text{jets}}\vec{p}^j_\T\right|
&&\text{for jets with} \quad p^j_\T>30 \gev\,,\; |\eta_j|<5\,.
\label{eq:htmissdef}
\end{align}
$H_\T$ characterizes the total visible hadronic activity and $\slashed{H}_\T$ the momentum 
imbalance in an event. The selection regions are categorized as summarized in table \ref{tab:MHT}.

\renewcommand{\arraystretch}{1.2}
\begin{table}[bt]
\centering
\begin{tabular}{crl}
\toprule
\multicolumn{1}{l}{bins} & \multicolumn{1}{l}{$H_\T\; [\gev\,]$} & $ \slashed{H}_\T \;[\gev\,]$\\
\midrule
1\,--\,4\phantom{1} & \phantom{1}500\,--\,800\phantom{1} & 200\,--\,300;~300\,--\,450;~450\,--\,600;~\textgreater\,600  \\
5\,--\,8\phantom{1} & \phantom{1}800\,--\,1000 & 200\,--\,300;~300\,--\,450;~450\,--\,600;~\textgreater\,600 \\
9\,--\,12 & 1000\,--\,1250 & 200\,--\,300;~300\,--\,450;~450\,--\,600;~\textgreater\,600 \\
13\,--\,15 & 1250\,--\,1500 & 200\,--\,300;~300\,--\,450;~\textgreater 450 \\
16\,--\,17 & \textgreater 1500 & 200\,--\,300;~\textgreater\,300  \\
\multicolumn{1}{c}{0} & \textgreater \phantom{1}500 &\textgreater\,200  \\
\bottomrule
\end{tabular}
\caption{Categorization of selection regions (bins) for the  $\slashed{H}_\T$ analysis \cite{Chatrchyan:2014lfa}.}
\label{tab:MHT}
\end{table}

Further cuts are $\left|\Delta\phi(\vec{p}_{j},\slashed{\vec{H}}_\T)\right|>0.5$ 
for the two hardest jets and $\left|\Delta\phi(\vec{p}_{j},\slashed{\vec{H}}_\T)\right|>0.3$
for the third hardest jet, as well as a veto against 
isolated electrons and muons with $p_\T>10\gev$. For details see \cite{Chatrchyan:2014lfa}.
We refer to this analysis in the following as the  $\slashed{H}_\T$ (MHT) analysis.\\

Another CMS analysis \cite{Chatrchyan:2013lya} is based on the variable $\alpha_\T$, which is a powerful 
variable for discrimination against QCD multijet background. It rejects multijet events without 
significant $\slashed{E}_\T$ \cite{Randall:2008rw,Khachatryan:2011tk}. 
For a dijet system, $\alpha_\T$ is defined as 
\begin{align}
&\alpha_\T=\frac{E_\T^{j_2}}{M_\T}\,,	&&	
M_\T=\sqrt{\left(\sum_{i=1}^2E_\T^{j_i}\right)^2-\left(\sum_{i=1}^2p_x^{j_i}\right)^2-\left(\sum_{i=1}^2p_y^{j_i}\right)^2},
\end{align}
where $j_2$ denotes the less energetic jet. $\alpha_\T$ is 0.5 for perfectly measured dijet events that 
are back-to-back in $\phi$. If the two jets are not back-to-back and recoil against a large 
$\slashed{E}_\T$, $\alpha_\T$ becomes larger than $0.5$. For suppression of mismeasured
QCD background, $\alpha_\T$ is required to be larger than 0.55. For events with more than two jets, 
a pseudo-dijet system is formed, such that the absolute $E_\T$ difference, denoted as $\Delta H_\T$, between the 
two pseudo-jets is minimized. In this case, the variable is generalized to
\begin{align}
    &\alpha_\T=  \frac{1}{2}\frac{H_\T-\Delta H_\T}{\sqrt{H_\T^2-\slashed{H}_\T^2}},
\end{align}
where $H_\T$ is the scalar sum of the transverse energies $E_\T$ of the jets, 
$H_\T=\sum_j E_\T^{j}$,\footnote{Note that this definition differs from the one used in the MHT analysis.}
and $\slashed{H}_\T$ is as defined in (\ref{eq:htmissdef}).
As we consider light flavor squark-squark production, we choose the signal region using 2--3 jets 
without a $b$-tagged jet. 
To maximize the sensitivity, the events are divided in eight bins based on $H_\T$: one bin in the range of 275--325\gev, one in the range of 325--375\gev,
five bins between 375--875\gev\ in steps of 100\gev, and an open bin $>875$\gev.
In the following we will refer to these eight bins as bin\,$1\dots 8$\, and to the combination of all bins as bin\,0. 
The selection criteria for jets in the first two bins differ from the others: for bin\,1 (bin\,2), 
$E_\T^j >37\gev$ $(43\gev)$ is required. In these same two bins, the two highest-$E_\T$  jets are required to have 
$E_\T^j >73\gev$ $(87\gev)$. For the other bins, the threshold 
for the two highest-$E_\T$ jets is $E_\T^j>100\gev$ each, and all jets are required to have $E_\T^j>50\gev$. In addition, for all 
bins jets are required to have $|\eta|<3$ ($|\eta|<2.5$ for the highest-$E_\T$ jet). For details see \cite{Chatrchyan:2013lya}. We refer to this analysis in the following as the $\alpha_\T$ analysis.

%----------------------------------------------------------------------------------------------------------------
\subsection{Event generation and parameter scan}
%----------------------------------------------------------------------------------------------------------------

In order to compute signal efficiencies, we performed a Monte Carlo event simulation.
We used \textsc{MadGraph}~5~\cite{Alwall:2011uj} to simulate the hard scattering of the squarks, 
whereafter \textsc{Pythia}~6~\cite{Sjostrand:2006za} was used for the decays of the squarks into 
neutralinos as well as for showering and hadronization. As the MHT analysis requires at least three
hard jets, in our model at least
one jet has to arise from initial state radiation. We therefore include up to one jet in the hard scattering
and performed an MLM matching \cite{alwall-2008-53} with initial state radiation from \textsc{Pythia}~6.
We chose a matching scale $Q_{\text{cut}}=46\gev$ and $p_\T^{\text{jet}}>30\gev$ in \textsc{MadGraph}~5. 
The gluino and squark\footnote{%
We assumed a pure bino for the computation of squark widths.}
widths were computed with \textsc{MadGraph}~5. The branching ratios of squark decays to the neutralino 
were set to 1. We used \textsc{Delphes}~3.0.11~\cite{deFavereau:2013fsa} with \textsc{Fastjet}~\cite{fastjet} 
for detector simulation using the standard CMS settings included in this version of \textsc{Delphes}, but changed the $b$-tag misidentification rate to 
0.01 \cite{Chatrchyan:2013lya,Chatrchyan:2012jua}.

We performed parameter scans in the mass region $m_{\s\chi^0_1}=100\dots1400\gev$ and 
$\msq=500\dots1500 \gev$ in steps of $100\gev$, with $\msq-m_{\s\chi^0_1}\geq 100\gev$, for 
two mass ratios $\mgo/\msq=2,4$. For both ratios, associated gluino and gluino pair production
are negligible.
We computed the efficiencies for MHT and $\alpha_\T$ for all sub-channels of first generation
squark production (as listed in section \ref{sec:production}), as well as for $\sq_1^{(*)}\sq_2^{(*)}+\sq_2^{(*)}\sq_2^{(*)}$, 
or production of at least one second generation squark (or anti-squark). Finally, as a reference
we computed the efficiencies for $pp\to \sq_\L^{}\sq_\L^*$ for the case $\mgo\to\infty$, which amounts to the simplified model $\Tinv$.

%===================================================================
\section{Results}
\label{sec:results}
%===================================================================
In this section, we show the deviations from the simplified model $\Tinv$ that arise from different production mechanisms, both at the efficiency level (see section \ref{section:efficiencies}) and at the level of squark and LSP mass limits (see section \ref{sec:limits}). We find that although there are some substantial differences at the efficiency level, the deviations in the mass limits based on current LHC data are rather moderate.

%----------------------------------------------------------------------------------------------------------------
\subsection{Efficiencies}
\label{section:efficiencies}
%----------------------------------------------------------------------------------------------------------------

\begin{figure}[h!]
\centering
\setlength{\unitlength}{1\textwidth}
\begin{picture}(0.85,1.32)
 \put(0.,-0.006){ 
  \put(0.0,0.025){\includegraphics[scale=0.61]{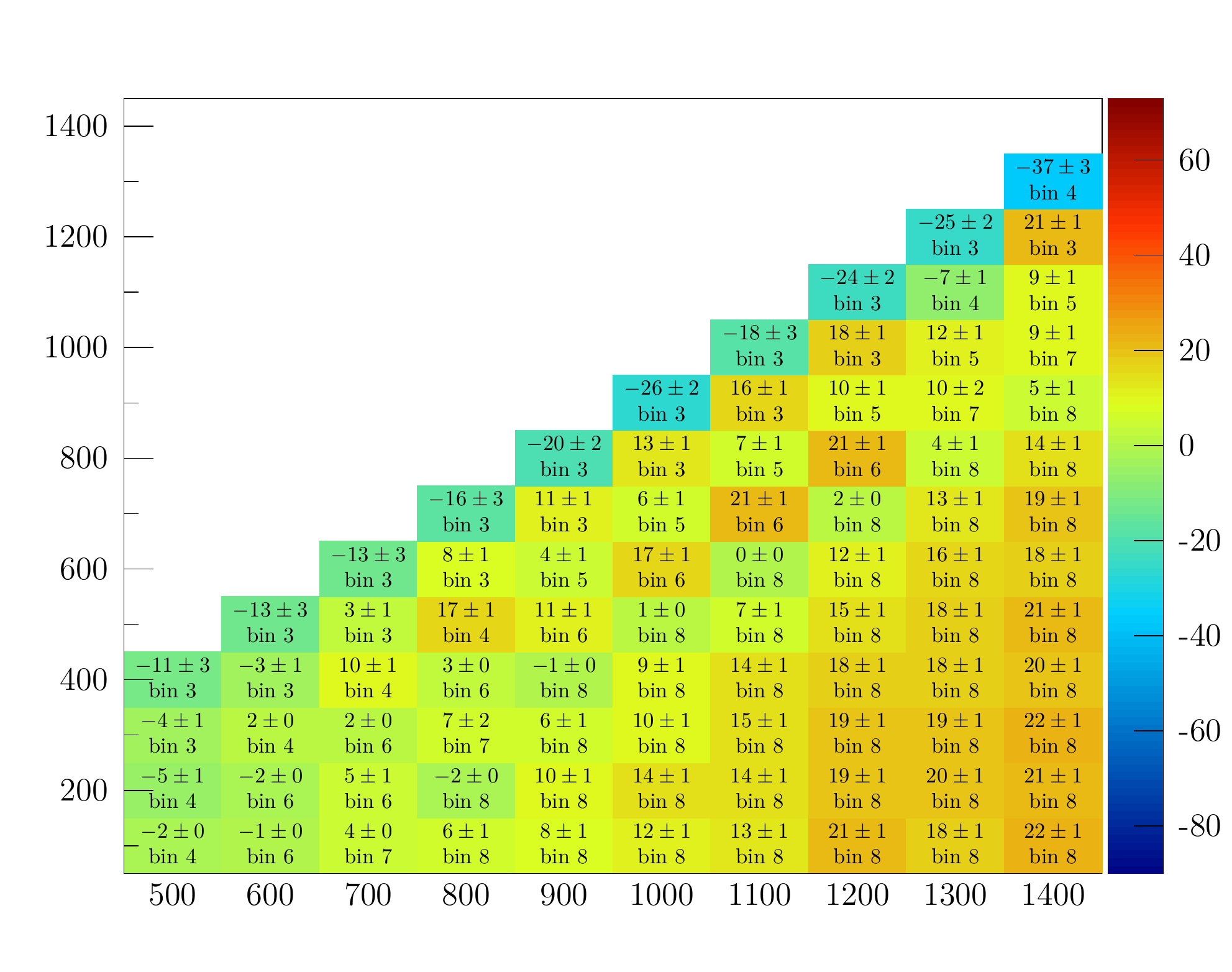}} 
  \put(0.12,0.56){$\sq_\L\sq_\L$, $\mgo/\msq=2$, $\alpha_{\text{T}}$}
  \put(0.38,0.02){\footnotesize $\msq\,[\gev\,]$}
  \put(-0.013,0.286){\rotatebox{90}{\footnotesize $m_{\s\chi^0_1}\,[\gev\,]$}}
  \put(0.835,0.525){\rotatebox{-90}{\footnotesize $\left(A\epsilon(\sq_\L\sq_\L)-A\epsilon(\Tinv)\right)/A\epsilon(\Tinv)\;[\,\%\,]$}}  
  }
\put(0,0.64){ 
  \put(0.0,0.025){\includegraphics[scale=0.61]{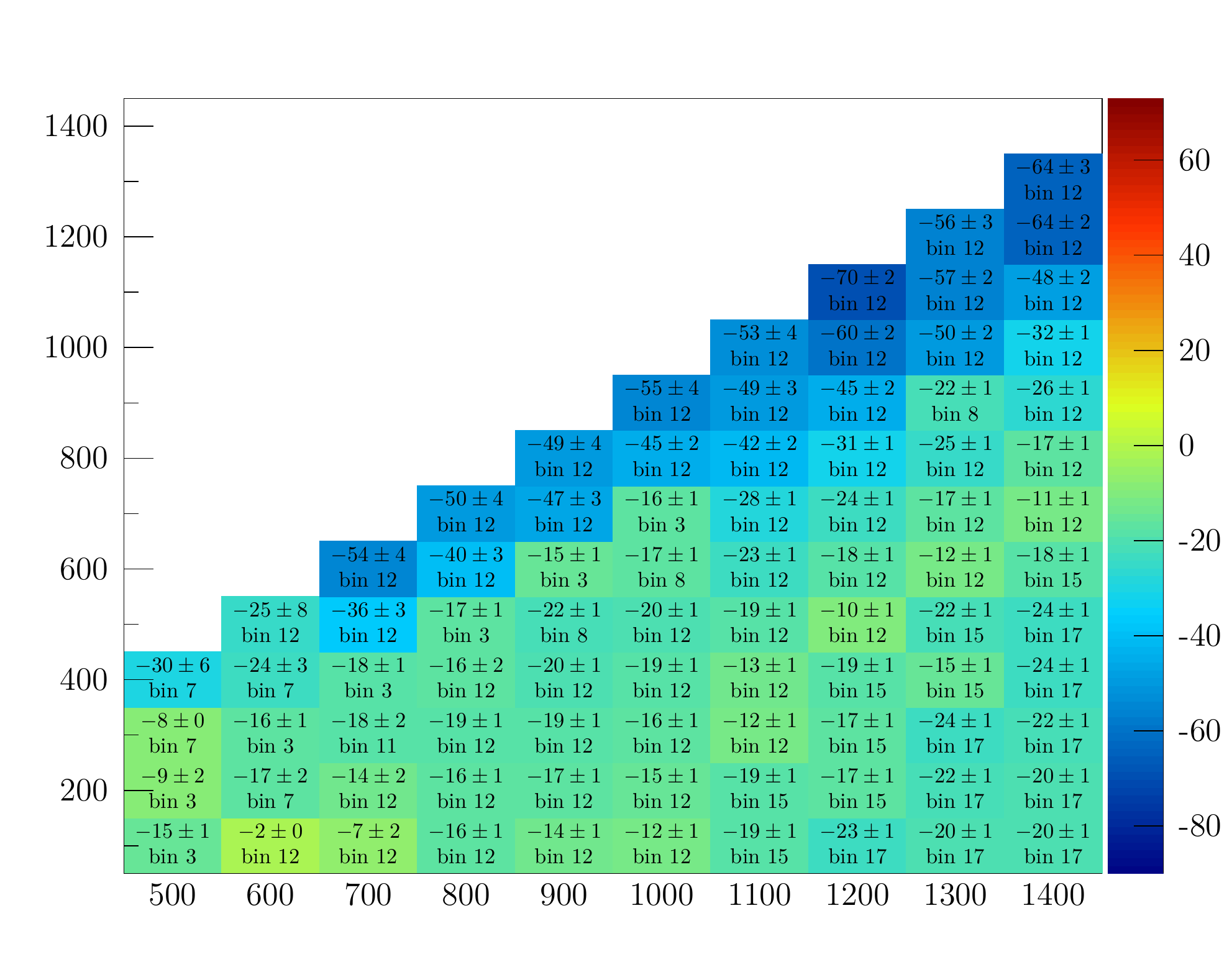}} 
  \put(0.12,0.56){$\sq_\L\sq_\L$, $\mgo/\msq=2$, MHT}
  \put(0.38,0.02){\footnotesize $\msq\,[\gev\,]$}
  \put(-0.013,0.286){\rotatebox{90}{\footnotesize $m_{\s\chi^0_1}\,[\gev\,]$}}
  \put(0.835,0.525){\rotatebox{-90}{\footnotesize $\left(A\epsilon(\sq_\L\sq_\L)-A\epsilon(\Tinv)\right)/A\epsilon(\Tinv)\;[\,\%\,]$}}  }
\end{picture}
\caption{Relative difference of signal efficiencies $A\epsilon$ for $\sq_\L^{}\sq_\L^{}$-production with $\mgo/\msq=2$ and 
$\sq_\L^{}\sq_\L^{\,*}$-production with $\mgo/\msq\to\infty$ ($\Tinv$) for the MHT analysis (upper panel) and
the $\alpha_{\text{T}}$-analysis (lower panel) at the $8\tev$ LHC\@. The relative difference is denoted by the color code and given 
in percent. The error corresponds to the Monte Carlo error from event generation. For each mass point we 
compare the efficiencies of the bin that yields the largest sensitivity within the $\Tinv$ model. The corresponding bin
is displayed below the relative difference.
}
\label{fig:accimportant}
\end{figure}

Comparing the signal acceptance\,$\times$\,efficiency $A\epsilon$ (simply called ``efficiency'' in the following) for $\Tinv$ and the other individual production 
mechanisms $M$ that contribute to $\Tmgo$, for individual bins\,$i$ we found very large relative deviations 
$$\frac{A\epsilon^i(M)-A\epsilon^i(\Tinv)}{A\epsilon^i(\Tinv)},$$
that were up to about 220\%. However, not all bins are equally relevant for setting exclusion limits.
In order to take this into account, we consider here the most sensitive bin only, which we define as 
the bin that yields the largest ratio $S^i/S^i_{95\%\,\text{C.L.}}(B)$,
where $S^i\propto A\epsilon^i$ is the expected number of signal events in bin\,$i$, while 
$S^i_{95\%\,\text{C.L.}}(B)$ is the corresponding required number of signal events that would provide a
95\%\,C.L. exclusion if the data would equal the background prediction $B$.
Of particular interest is the production mechanism $pp\rightarrow \widetilde{q}_\L^{}\widetilde{q}_\L^{},\sq_\R^{}\sq_\R^{}$,
as it is the dominant contribution (see section \ref{sec:production}).
The deviations in the efficiencies obtained for the production mode $pp\to\sq_\L^{}\sq_\L^{}$ from those obtained for $\Tinv$ are shown in figure 
\ref{fig:accimportant} for $\mgo/\msq=2$.

\subsubsection*{MHT analysis}
An overall feature for the MHT analysis is that the $A \epsilon$ for $M=\widetilde{q}_\L^{}\widetilde{q}_\L^{}$ are smaller than those for $\Tinv$. The deviations range from up to $- 70\%$ in the region where $m_{\widetilde{q}}-m_{\s\chi^0_1}\approx 100\gev$, to a few percent for small LSP and squark masses. 
For most masses bin\,12 or bin\,17 (for large squark and small LSP masses) are the most sensitive bins.
Another important production mode is  $\widetilde{q}_\L^{}\widetilde{q}_\R^*$ which contributes significantly to the total cross section for $m_{\widetilde{g}}/m_{\widetilde{q}}\approx 2-5$. For $m_{\widetilde{g}}/m_{\widetilde{q}}=2$, the deviations are largest for $m_{\widetilde{q}}-m_{\s\chi^0_1}\approx 100$\,GeV and reach up to $-80$\%. The deviations decrease to $-5$\% in the region $m_{\widetilde{q}}-m_{\s\chi^0_1}\approx 700$\,GeV and increase again up to $-20$\% for large squark and small LSP masses. For gluino masses $m_{\widetilde{g}}/m_{\widetilde{q}}=4$, the general features remain the same, but the absolute values of the deviations increase by a few percentage points.  All deviations are negative, meaning that the $A\epsilon$ for $\Tinv$ is larger then for $\widetilde{q}_\L^{}\widetilde{q}_\R^*$.
For $\widetilde{q}_\L^{}\widetilde{q}_\R^{}$ we found deviations as large as 220\%. However, the contribution of $\widetilde{q}_\L^{}\widetilde{q}_\R^{}$ to the total cross section is completely negligible for $\mgo/\msq=4$ and rather small for $\mgo/\msq=2$.

%=====================
%    \                                           |
%      \                                         |
%        \                                       |
%                                      \         |
%                                        \       |
%                                          \     |
%=====================

\clearpage

\subsubsection*{$\alpha_\T$ analysis}
The $\alpha_\T$ analysis turns out to be more robust towards the different production and gluino mass scenarios. The efficiencies for $\widetilde{q}_\L^{}\widetilde{q}_\L^{}$ are mainly larger than for $\Tinv$ (with the exception of small squark/LSP masses and mass differences). The deviations are in the range of $+20$ to $-40$\%. The most sensitive bin for large squark and small neutralino masses is bin\,8.
The deviations of the production mode $\widetilde{q}_\L^{}\widetilde{q}_\L^*$ from $\Tinv$ are around a few percent. The largest deviations occur for small squark and LSP masses with around $-10$\%. For larger $m_{\widetilde{g}}/m_{\widetilde{q}}=4$, the deviations become even smaller, as expected. The  $A\epsilon$ for  $\widetilde{q}_\L^{}\widetilde{q}_\R^*$ are larger than for $\Tinv$. 
The largest deviations are $\approx 30$\% for large squark masses. 

%----------------------------------------------------------------------------------------------------------------
\subsection{Exclusion limits}
\label{sec:limits}
%----------------------------------------------------------------------------------------------------------------

Our final goal is to examine the effect of the different production mechanisms and gluino masses on the search sensitivity.
From the efficiencies we derived 95\% C.L.\ exclusion limits in the $\msq$-$m_{\s\chi^0_1}$ plane for 
$\mgo/\msq=2,4$, for both the simplified model $\Tinv$ and the complete supersymmetric model $\Tmgo$.
In order to obtain the efficiencies for $\Tmgo$ from the individual production channels, we compute the 
weighted mean of the efficiencies, where we use NLO cross sections computed with 
\textsc{Prospino\,2}~\cite{Beenakker:1996ch} for the weights. 
For the determination of the most sensitive bin for $\Tmgo$ and $\Tinv$, we
included bin\,0 (the sum of all bins of the analysis). The most sensitive bin is determined, as before, on the basis of background expectation only. 
To obtain the exclusion limits, we compute
\begin{align}
    \mu^{-1}=\left(\frac{(A\epsilon)^i\times \int\! \mathcal{L} \times\sigma_{\text{tot}} }{S^i_{95\%\text{CL}}(\text{data})}\right)_{i\,=\,\text{most sensitive bin}},
\end{align}
for each point in the $\msq$-$m_{\s\chi^0_1}$ plane. Here, $S^i_{95\%\text{CL}}(\text{data})$ is the 
required number of signal events allowing for a 95\% C.L.\ exclusion in the presence of the measured
number of events (data). The total cross section $\sigma_{\text{tot}}$ was computed from
\textsc{Prospino\,2}~\cite{Beenakker:1996ch} and multiplied with NLL $K$-factors from 
\textsc{NLLfast}~\cite{Beenakker:2009ha}.\footnote{%
    For points with $\mgo>2500\gev$, we used the respective $K$-factor for $\mgo=2500\gev$, as larger 
values are not provided in \textsc{NLLfast}. \label{ftn:NLLgrid}}

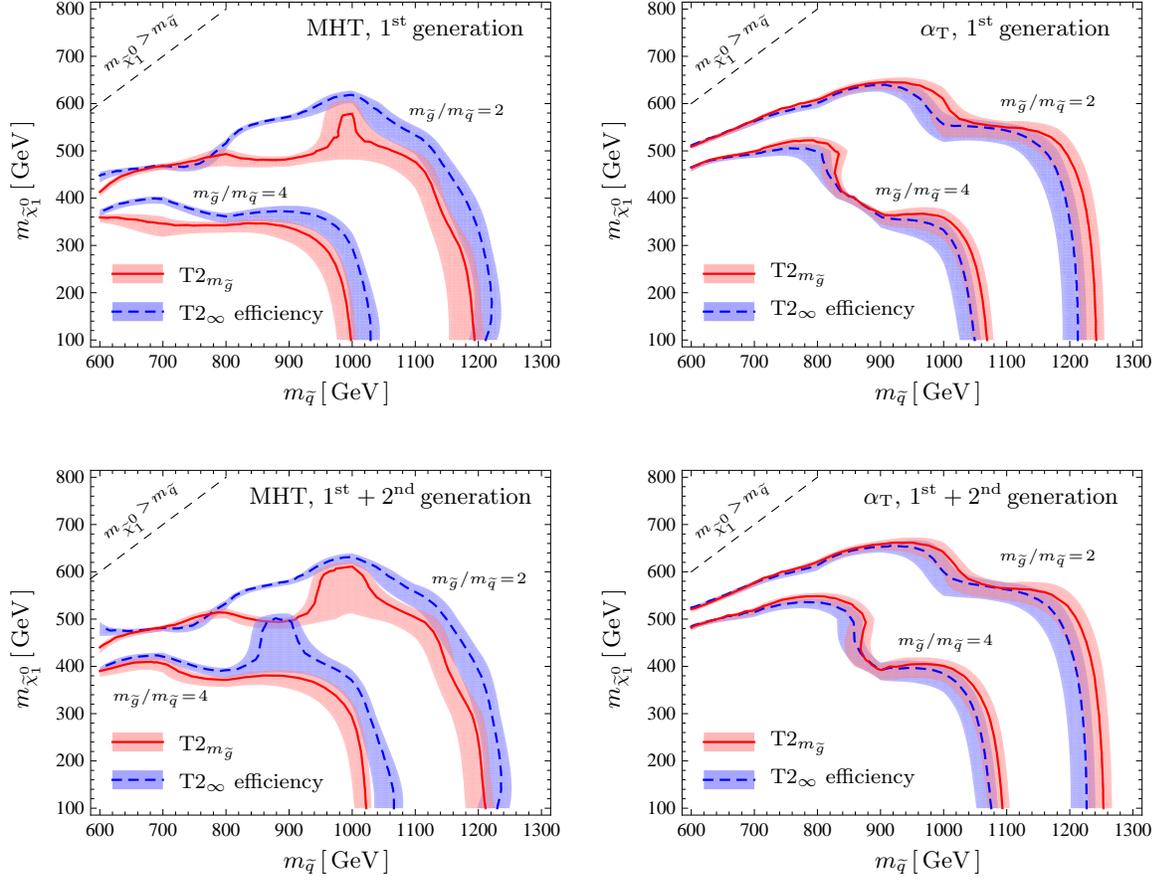
\begin{figure}[t]
\centering
\setlength{\unitlength}{1\textwidth}
\begin{picture}(0.937,0.778)
 \put(0.52,0.42){ 
  \put(0.0,0.025){\includegraphics[scale=1.1]{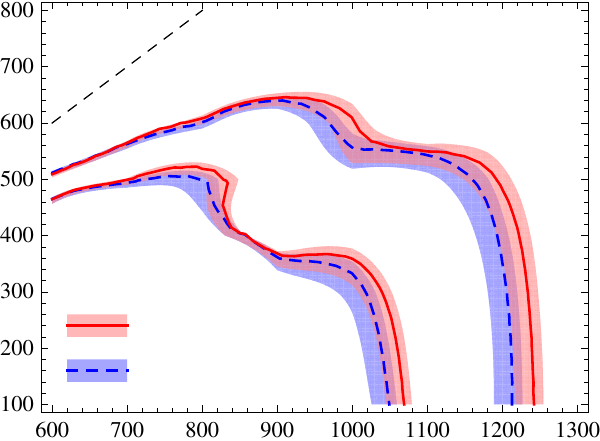}} 
  \put(0.24,0.32){\footnotesize $\alpha_{\text{T}}$, $1^{\text{st}}$\,generation}
  \put(0.2,0.0){\footnotesize $\msq\,[\gev\,]$}
  \put(-0.04,0.14){\rotatebox{90}{\footnotesize $m_{\s\chi^0_1}\,[\gev\,]$}}
  \put(0.31,0.258){\tiny $\mgo/\msq\!=\!2$}
  \put(0.2,0.18){\tiny $\mgo/\msq\!=\!4$}
  %key
  \put(0.108,0.107){\scriptsize $\Tmgo$}
  \put(0.108,0.074){\scriptsize $\Tinv$ efficiency}
  \put(0.04,0.288){\rotatebox{37.6}{\tiny $m_{\s\chi^0_1}\!>\!\msq$}}
  }
\put(0,0.42){ 
  \put(0.0,0.025){\includegraphics[scale=1.1]{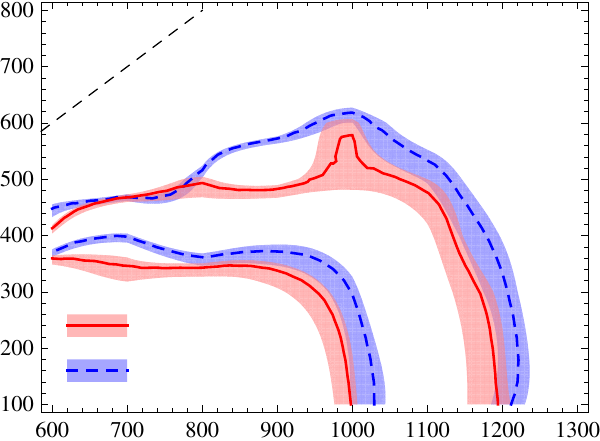}} 
  \put(0.22,0.32){\footnotesize MHT, $1^{\text{st}}$\,generation}
  \put(0.2,0.0){\footnotesize $\msq\,[\gev\,]$}
  \put(-0.04,0.14){\rotatebox{90}{\footnotesize $m_{\s\chi^0_1}\,[\gev\,]$}}
  \put(0.31,0.25){\tiny $\mgo/\msq\!=\!2$}
  \put(0.12,0.177){\tiny $\mgo/\msq\!=\!4$}
   %key
  \put(0.108,0.103){\scriptsize $\Tmgo$}
  \put(0.108,0.0697){\scriptsize $\Tinv$ efficiency}
  \put(0.04,0.288){\rotatebox{37.6}{\tiny $m_{\s\chi^0_1}\!>\!\msq$}}
  }
 \put(0.52,0.008){ 
  \put(0.0,0.025){\includegraphics[scale=1.1]{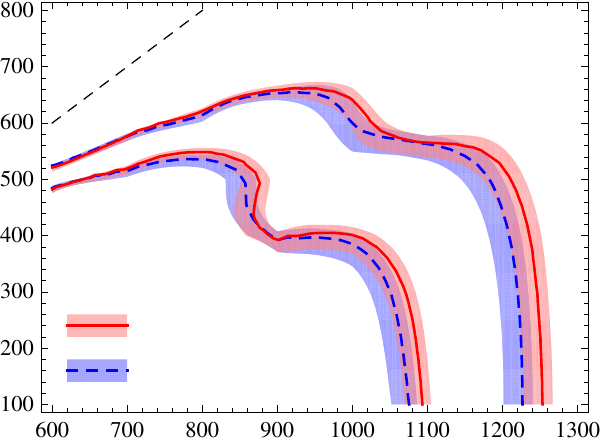}} 
  \put(0.19,0.32){\footnotesize $\alpha_{\text{T}}$, $1^{\text{st}}+2^{\text{nd}}$\,generation}
  \put(0.2,0.0){\footnotesize $\msq\,[\gev\,]$}
  \put(-0.04,0.14){\rotatebox{90}{\footnotesize $m_{\s\chi^0_1}\,[\gev\,]$}}
  \put(0.31,0.27){\tiny $\mgo/\msq\!=\!2$}
  \put(0.22,0.196){\tiny $\mgo/\msq\!=\!4$}
  %key
  \put(0.108,0.107){\scriptsize $\Tmgo$}
  \put(0.108,0.074){\scriptsize $\Tinv$ efficiency}
  \put(0.04,0.288){\rotatebox{37.6}{\tiny $m_{\s\chi^0_1}\!>\!\msq$}}
  }
\put(0,0.008){ 
  \put(0.0,0.025){\includegraphics[scale=1.1]{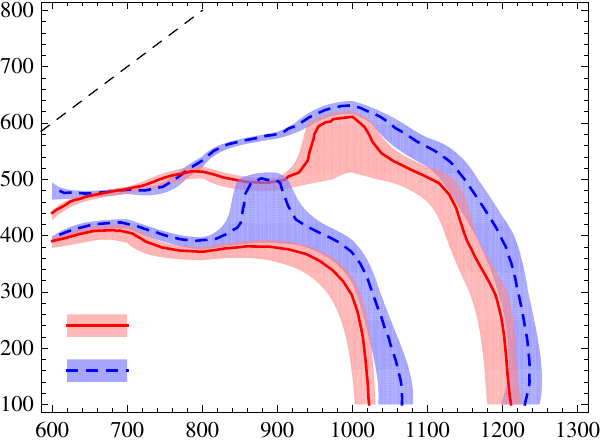}} 
  \put(0.17,0.32){\footnotesize MHT, $1^{\text{st}}+2^{\text{nd}}$\,generation}
  \put(0.2,0.0){\footnotesize $\msq\,[\gev\,]$}
  \put(-0.04,0.14){\rotatebox{90}{\footnotesize $m_{\s\chi^0_1}\,[\gev\,]$}}
  \put(0.33,0.25){\tiny $\mgo/\msq\!=\!2$}
  \put(0.05,0.147){\tiny $\mgo/\msq\!=\!4$}
   %key
  \put(0.108,0.103){\scriptsize $\Tmgo$}
  \put(0.108,0.0697){\scriptsize $\Tinv$ efficiency}
  \put(0.04,0.288){\rotatebox{37.6}{\tiny $m_{\s\chi^0_1}\!>\!\msq$}}
  }
\end{picture}
\caption{95\% C.L. exclusion limits derived from the full $\Tmgo$ model (red, solid curves) and
from the efficiencies for the $\Tinv$ simplified model (blue, dashed curves). The shaded regions around
the curves denote the uncertainties due to scale variation ($\mu=\msq/2,2\msq$). The $\alpha_{\text{T}}$
and MHT analyses are based on an integrated luminosity of 11.7\,fb$^{-1}$ and 19.5\,fb$^{-1}$, respectively,
collected at the $8\tev$ LHC\@.
}
\label{fig:exclim}
\end{figure}

Figure \ref{fig:exclim} shows the 95\% C.L. exclusion limits -- the contours $\mu\!=\!1$ -- 
for the MHT and $\aT$ analyses for 
first generation squarks only as well as first and second generation squarks,
both for $\mgo/\msq=2,4$. The shaded bands around the exclusion 
limits denote the uncertainties from scale variation, which we took to be $\mu = \msq/2, 2\msq$.\footnote{%
We show here the scale variation at NLO\@. The information was not available at NLL
accuracy for the complete parameter space considered (see footnote \ref{ftn:NLLgrid}).
However, although scale uncertainties at NLL are smaller than at NLO, additional 
uncertainties from the parton distribution functions and $\alpha_\text{s}$ should, in principle, be taken into account. 
Hence, we expect the presented uncertainties to give a reasonable estimate of the over-all theoretical uncertainty.}

We observe the following results: First, for both $\mgo/\msq=2$ and $4$, the deviations in the 
exclusion limits derived from the efficiencies taken from $\Tinv$ and the full supersymmetric model $\Tmgo$ 
are of the order of the theoretical uncertainty on the cross section normalization for large regions 
in parameter space. Second, whereas for the $\aT$ analysis the uncertainty bands 
overlap throughout the exclusion limit, there are deviations in the MHT analysis in the 
region $\mne\gtrsim\msq/2$. The upper limit on $\mne$ for $\msq<1\tev$ is
very sensitive to small changes in the efficiencies. The $\aT$ analysis is
much less sensitive to the actual production mode and thus less model-dependent.  
Third, while for the MHT analysis the limits from $\Tinv$ overestimate 
the exclusion limits for most of the parameter space, the $\aT$
analysis $\Tinv$  limits stay conservative over the complete parameter
space. 

\vspace*{5mm}

%=====================
%    \                                           |
%      \                                         |
%        \                                       |
%                                      \         |
%                                        \       |
%                                          \     |
%=====================

%\clearpage

%===================================================================
\section{Conclusion}\label{sec:conclusion}
%===================================================================

In this study we investigated the validity of a simplified model description of 
squark production at the LHC\@. Concentrating on the T2 topology, $pp\to\sq^{(*)}\sq^{(*)}\to q q\s\chi^0_1\s\chi^0_1$,
we examined the effect of a varying gluino mass and thus of varying dominant production channels. 
We found that the often used limiting case of a simplified model with a decoupled gluino (here denoted as $\Tinv$) 
is a priori not a good description unless $\mgo \gg 10\,\msq$. For most of the relevant parameter
space for $\mgo<10\, \msq$, squark production is dominated by the production
channels $pp\to\sq_\L^{}\sq_\L^{},\sq_\R^{}\sq_\R^{}$; this in contrast to the case of a decoupled gluino, where only same-flavor
and same-chirality squark-anti-squark production is present. We computed the signal efficiencies 
for two all-hadronic analyses performed by CMS: one based on missing transverse energy (MHT), and
one based on the $\aT$ variable. We found a larger sensitivity on the production channel in the
MHT analysis. For the most sensitive bin in the analyses, we found relative deviations between the 
efficiencies from $\Tinv$ and the dominant production mode ($pp\to\sq_\L^{}\sq_\L^{},\sq_\R^{}\sq_\R^{}$) of up to
70\% for the MHT analysis and up to 40\% for the $\aT$ analysis. However, both maximal 
differences were found in the region of large $\msq$ and small mass splittings $\msq-\mne$,
which is far beyond the exclusion limits that could be derived from the $8\tev$ LHC run. Hence,
we found little deviation between the derived mass exclusion limits from the $\Tinv$ simplified model efficiencies and those of the
full supersymmetric model $\Tmgo$. In particular, we found that limits derived from the
$\aT$ analysis are much less sensitive to the production mode. For the $\aT$ analysis,
the limits from $\Tinv$ provide conservative estimates for the limits within the full model. In contrast, $\Tinv$ tends to overestimate the limits in the case of the MHT analysis.

We showed our results for $\mgo/\msq=2,4$. For $\mgo/\msq>4$ and $\mgo/\msq<2$, the deviations
in the efficiencies tend to become smaller. For the former case, the same-flavor
and same-chirality squark-anti-squark production becomes more important, which is much less 
dependent on the gluino mass and hence resembles $\Tinv$ for large 
$\mgo/\msq$. For $\mgo/\msq<2$, the production channel $\sq_\L^{}\sq_\R^{}$ becomes important.
However, the differences in the efficiencies between $\sq_\L^{}\sq_\R^{}$ and $\Tinv$ tend to decrease
with decreasing $\mgo/\msq$. In particular, we found smaller differences than for $\sq_\L^{}\sq_\L^{}$
with $\mgo/\msq=1$. However, for $\mgo/\msq\lesssim1$ associated squark-gluino production becomes
dominant. In this part of parameter space, the gluino mass has to be taken into account as an additional 
parameter in the simplified model analysis. 

To conclude, the simplified model $\Tinv$ is a  reliable tool to interpret the current hadronic jets plus missing energy 
searches at the LHC in a more model-independent way, in particular for analyses based on the variable $\aT$. Larger differences between simplified 
model and complete supersymmetric model interpretations could, however, arise in future LHC searches for heavier squarks, depending in detail 
on the experimental cuts and the parameter space probed. 

\section*{Acknowledgements}

We would like to thank Christian Autermann, Lutz Feld and Wolfgang Waltenberger for useful discussions and suggestions. 
We are grateful to Wolfgang and to the Institute of High Energy Physics (HEPHY) for their hospitality during various visits 
to Vienna. MK is grateful to SLAC and Stanford University for their hospitality during his sabbatical stay. This work was supported by the Deutsche 
Forschungsgemeinschaft through the graduate school ``Particle and Astroparticle Physics in the Light of the LHC'' and through 
the collaborative research centre TTR9 ``Computational Particle Physics'',  by the German Federal Ministry of Education and Research BMBF, 
and by the U.S.\ Department of Energy under contract DE-AC02-76SF00515.

%\clearpage

\addcontentsline{toc}{section}{References}
\bibliographystyle{utphys}
\bibliography{SMSref}

%%%%%%%%%%%%%%%%%%%%%%%%%%%%%%%%%%%%%%%%%%%%%%%%%%%%%%%
\end{document}